\documentclass[conference]{IEEEtran}
\IEEEoverridecommandlockouts
\usepackage{cite}
\usepackage{amsmath,amssymb,amsfonts}
\usepackage{algorithmic}
\usepackage{graphicx}
\usepackage{textcomp}
\usepackage[table]{xcolor}
\usepackage{xcolor}
\def\BibTeX{{\rm B\kern-.05em{\sc i\kern-.025em b}\kern-.08em
    T\kern-.1667em\lower.7ex\hbox{E}\kern-.125emX}}
\begin{document}

\title{\huge \textbf{Federated TON\_IoT Windows Datasets for Evaluating AI-based  Security Applications\\ \thanks{This work was funded by Australian Research Data Commons (\textit{RG192500}) and UNSW Canberra (\textit{PS51776}). Free use of the TON\_IoT datasets for academic purposes is hereby granted in perpetuity. Use for commercial purposes is allowable after asking the author, Dr Nour Moustafa, who has asserted his right under the copyright.}  }}

\author{\IEEEauthorblockN{Nour Moustafa\IEEEauthorrefmark{1}, Marwa Keshk\IEEEauthorrefmark{2}, Essam Debie\IEEEauthorrefmark{3} and Helge Janicke\IEEEauthorrefmark{4}}
\IEEEauthorblockA{School of Engineering and Information Technology, University of New South Wales, Canberra, Australia \IEEEauthorrefmark{1}\IEEEauthorrefmark{2}\IEEEauthorrefmark{3}\\
Cyber Security Cooperative Research Centre, Edith Cowan University, Perth, Australia\IEEEauthorrefmark{4}\\}
Email: \IEEEauthorrefmark{1}nour.moustafa@unsw.edu.au,
\IEEEauthorrefmark{2}marwa.hassan@student.unsw.edu.au, \\
\IEEEauthorrefmark{3}e.debie@unsw.edu.au,
\IEEEauthorrefmark{4}helge.janicke@cybersecuritycrc.org.au}

\setlength{\columnsep}{0.2in}
\maketitle

\begin{abstract}
 Existing cyber security solutions have been basically developed using knowledge-based models that often cannot trigger new cyber-attack families. With the boom of Artificial Intelligence (AI), especially Deep Learning (DL) algorithms, those security solutions have been plugged-in with AI models to discover, trace, mitigate or respond to incidents of new security events. The algorithms demand a large number of heterogeneous data sources to train and validate new security systems. This paper presents the description of new datasets, the so-called \textbf{ToN\_IoT}, which involve federated data sources collected from \textbf{T}elemetry datasets of \textbf{IoT} services, \textbf{O}perating system datasets of Windows and Linux, and datasets of \textbf{N}etwork traffic. The paper introduces the testbed and description of TON\_IoT datasets for Windows operating systems.  The testbed was implemented in three layers: edge, fog and cloud. The edge layer involves IoT and network devices, the fog layer contains virtual machines and gateways, and the cloud layer involves cloud services, such as data analytics, linked to the other two layers. These layers were dynamically managed using the platforms of software-Defined Network (SDN) and Network-Function Virtualization (NFV) using the VMware NSX and vCloud NFV platform. The Windows datasets  were collected from audit traces of memories, processors, networks, processes and hard disks. The datasets would be used to evaluate various AI-based cyber security solutions, including intrusion detection, threat intelligence and hunting, privacy preservation and digital forensics. This is because the datasets have a wide range of recent normal and attack features and observations, as well as authentic ground truth events. The datasets can be publicly accessed from this link \cite{ton-data}.    
\end{abstract}


\section{Introduction}
\label{intro}

Through technological advancement in the ICT sector, the Internet of Things (IoT) is becoming an important aspect of society to offer automated services to users and organizations. As IoT systems are assigned ever more important tasks, often handling sensitive data or critical infrastructure, it is evident that any disruption caused by malicious actors will result in tremendous problems such as loss of businesses or human lives \cite{moustafa2019systemic}. As a result, shielding these technologies through cyber security and ensuring their stability is imperative.  Due to the reliance on the Internet, there are many security vulnerabilities in IoT networks and their operating systems, such as Windows and Linux, where attackers could exploit systems to gain access and steal important information from them \cite{ho2016smart,seralathan2018iot,koroniotis2019forensics}. As a result, tremendous cyber attacks are expected to target IoT systems through operating systems because the IoT systems are easy to use and have been designed to be continuously on-line, providing a reliable platform from which to launch complex hacking events.

The Operating System (OS) is the core of computer systems and their networks that manage hardware underpinning and provide low-level services to high-level programs. Although different types of OSs have been developed over the years, in this paper we focus on Windows OS, especially Windows 7 and Windows 10. The Windows OS is a family of proprietary operating systems, owned and maintained by Microsoft \cite{yosifovich2017windows}. Windows OSs  available in various computer systems include Windows 7, Windows 10, Windows Server, Windows NT, Windows IoT, and Windows Mobile. Windows is the world's most used OS \cite{majeed2017forensic,aladdin2018effects}, and has often been the target of cyber-attacks like ransomware and malware \cite{akkas2017malware,kim2017certified,zimba2018multi} with significant consequences. 

Investigating cyber threats targeting Windows OS and developing new Artificial Intelligence (AI)-based cyber security solutions, including intrusion detection, threat intelligence and hunting, privacy preservation and digital forensics, have a great interest to improve system security. Multiple cyber threats have encouraged the design and development of various cyber defensive mechanisms, with specialized techniques that focus on specific device subsystems \cite{liao2013intrusion,moustafa2017big}. Intrusion Detection Systems (IDS) are one particular family of defensive mechanisms that have been the focus of much research in recent years. \cite{moustafa2017collaborative}. IDSs can be categorized into two types: 1) Host IDSs (HIDSs) are installed and monitor a host machine; and 2) Network IDSs are placed in strategic locations of a network, monitoring inbound and outbound traffic. Depending on their inner mechanism, IDS are characterized as either being anomaly-based, flagging unusual behavior, or signature-based, detecting known patterns of misuse \cite{moustafa2017big}.

In order to evaluate IDSs and AI-based security solutions, it is vital to use high-quality data that realistically represent current behavioral scenarios, including both attacks and legitimate events. Existing datasets that can be used for the development of a HIDS face several issues. To begin with, research has been focused on Linux OS, with most of the available datasets being derived from Linux-based testbeds that are different from audit traces of Windows OSs. Furthermore, existing datasets were generated from testbeds that did not incorporate IoT scenarios, where multiple lightweight devices communicate with each other to generate diverse normal patterns from conventional network systems  \cite{creech2013generation,sharafaldin2018toward}. This is a considerable weakness, as it is exceedingly common to have smart devices active in a network, whether it is public or private. In addition, most datasets were designed to be used for HIDS development, focus entirely on API calls and system calls  \cite{creech2013generation}, neglecting data derived from other subsystems, such as memory, processor, process and hard disk, which would cause that the defensive mechanisms often ignore complex attacks. Finally, most of the existing  datasets suffer from low credibility because they do not have the ground truth of security events or the provided data analysis is poor \cite{lippmann2000evaluating,creech2013generation}.

This paper addresses the issues above by introducing new Windows datasets that comprise new features extracted from the audit traces of memory, processor, process and hard disk in a new IoT network architecture. The testbed was deployed in three tiers, edge, fog and cloud. The edge tier includes IoT and network devices, the fog layer involves virtual machines and gateways, and the cloud tier comprises cloud services such as data analytics and visualization linked to the other tiers. These tiers were elastically managed using the technologies of software-Defined Network (SDN) and Network-Function Virtualization (NFV) using the VMware NSX \cite{SDN} and vCloud NFV \cite{NFV} platforms. While configuring and deploying the testbed, normal and attack events were executed to gather labeled data samples based on an authentic ground truth table for evaluating the performances of new cyber security applications.

The main contributions of this paper are as follows:
\begin{itemize}
  
  \item 	A new IoT testbed architecture is proposed for concurrently collecting federated data from heterogeneous sources, along with recent normal and malicious scenarios. 
  \item 	Software-Defined Network connectivity and Network Function Virtualization are proposed in the testbed to offer dynamic virtualization services in the proposed testbed.
  \item 	 New Windows-derived datasets that incorporate a wide range of recent legitimate and adversarial events were generated from the activities of hard disks, memories, processes, processors and network-level, for the validation of new AI-based cyber security solutions such as HIDSs.

\end{itemize}

The structure of this paper is as follows. Section \ref{background} describes the background and related work on IDSs and their datasets. Section \ref{testbed} illustrates the testbed designed for generating the TON\_IoT datasets. Hacking scenarios used in the datasets are explained in Section \ref{hacking}. Following that, the features and statistics of Windows 7 and 10 datasets are explained in Section \ref{expereiments}. Finally,  Section \ref{conclusion} presents concluding remarks of the work and future research directions.

\section{Background and Related work}
\label{background}

This section provides background information about IDSs and their previous studies. Several existing datasets are also explained for the purpose of constructing and evaluating IDSs in Windows environments. 

\subsection{Intrusion Detection Systems (IDSs)}

IDSs can be classified into two main categories, network-based and host-based IDS \cite{kozushko2003intrusion,liao2013intrusion}. Network-based IDSs (NIDSs) are platforms placed in a network's strategic points, such as gateways, to monitor traffic sent to and from the internal network for any suspicious patterns. Host-based IDSs (HIDS) are software agents to track host computers, relying on data, such as system calls and logs, to detect unauthorized activities.

An IDS can also be categorized on the basis of the detection method used in three categories: signature-based detection, anomaly-based detection, and hybrid of the first two. A signature-based detection \cite{freeman2002host} utilizes pre-defined patterns, known as signatures, which can be used to identify attacks. The idea of using signatures to detect intrusions is that a database of patterns is maintained and regularly updated to match and detect attack events. A signature-based HIDS monitors a host’s state by scanning various logs, memory dumps and network traffic generated or received by the host. They produce a high detection rate for known attacks and produce results at high speeds. However, signature-based solutions are hindered by even small changes to the known attack patterns (a common defensive technique employed by attackers). In addition, as they rely on known patterns, signature-based techniques are unable to detect unknown attack patterns, also known as zero-day exploits \cite{jose2018survey}.

An Anomaly-based HIDS, on the other hand, establishes a profile of legitimate usage, then flagging any activities that deviate from the profile as an intrusion \cite{jose2018survey}. As the decision engine, they make use of the machine and deep learning models, which are trained on collected data and learn to detect legitimate behavior. Because anomaly-based HIDSs fit normal behavior, they are capable of detecting zero-day attacks. In addition, they are capable of detecting mutations in attack patterns, a process often employed by malware to circumvent signature-based IDSs \cite{moustafa2019outlier}.
 
Although relying on the machine learning and deep learning models render anomaly-based HIDS more versatile compared to signature-based ones, they are also more computationally intensive \cite{bijone2016survey,jose2018survey}. Additionally, signature-based techniques could outperform anomaly-based models, when tasked with detecting known patterns. The higher false alarm rates produced by anomaly-based techniques would come as a result of the generalization that machine learning techniques rely on, to make predictions on unknown data points \cite{moustafa2018anomaly}. Building effective cyber security solutions, for example, intrusion detection, privacy preservation, threat hunting and intelligence and digital forensics, require proper datasets that include a wide range of recent legitimate and attack events to train and validate the security solutions effectively.
 
 \subsection{Datasets used for validating defensive mechanisms}
 
Acquiring a reliable dataset, which includes contemporary attack and normal scenarios, is an important first step towards designing, developing and validating effective AI-enabled defensive applications such as intrusion detection, privacy preservation, threat hunting and intelligence and digital forensics. Several datasets have been proposed in the literature which can be employed in IDS development, with particular attention given on operating system datasets \cite{koroniotis2019towards,moustafa2015unsw}. The most commonly used datasets are briefly discussed as follows:

 \begin{itemize}
 \item 	The DARPA 98 and KDD-99 dataset \cite{lippmann2000evaluating} -- is the first dataset intended for IDS research. The entire dataset includes 4GB of raw pcap files and was derived from a testbed that was modeled by the MIT Lincoln Laboratory. A number of attacks targeted Unix machines located in the testbed’s network, within a period of seven weeks. Although the dataset is labeled and widely used to date, the attack scenarios are out of date since they were generated in 1998, and do not reflect new normal and abnormal behaviors.

 \item 	The ADFA-LD dataset \cite{creech2013generation} -- was produced by the cyber-security centre at the University of New South Wales at ADFA. It is a Linux-based dataset generated from a testbed where an Ubuntu OS machine was targeted by a range of diverse attacks, such as password brute force, privilege escalation and Meterpeter-generated payloads. After scanning the Ubuntu machine, three datasets were produced  raw system call data, the first for training on normal data, the second for validation on normal data and the last set comprised of attack data.
 
 \item The NGIDS-DS dataset \cite{haider2017generating} -- was created by the cyber-security centre at the University of New South Wales at ADFA. The dataset has a combination of network traces, and host-based data, produced by utilizing the IXIA perfect storm tool and scanning the cyber-range set-up for a period of four days, producing more than ninety million records of network and host data. In the testbed, two Ubuntu OS machines were incorporated, one of which collected network traffic and the other recorded process-related data with a particular focus on system calls.

 \end{itemize}
 
 Although the aforementioned datasets were derived from Linux OS systems, a few datasets derived from Windows machines have also been proposed, as explained below.
 
 \begin{itemize}
 
  \item The SSENet-2014 dataset \cite{bhattacharya2014ssenet} -- was produced by extending the SSENet-2011 dataset. This dataset was generated by attacking a vulnerable Windows server, resulting in 28 attributes in total, with 9 basics, 9 network traffic and 10 host attributes. Although the statistics of the dataset were not provided, the dataset’s attributes match those of KDD-99.
  
  \item The AWSCTD dataset \cite{vceponis2018towards} -- was designed as an alternative to ADFA-WD in order to provide a current Windows-based dataset. To generate the dataset, Windows 7 guest machines were infected with a total of 10276 malware samples, producing 112.56 million audit traces. 
  
  \item The ADFA-WD dataset \cite{haider2016windows} -- was created by the cyber-security centre at the University of New South Wales, in an attempt to provide a reliable windows-based dataset for IDS. In the data-generation phase, a Windows XP machine was targeted by several attacks, including a zero-day attack and system calls in the form of a dynamic link library and called function addresses were recorded.

 \end{itemize}

While the research studies focused on developing Linux-based datasets, the limited number of Windows datasets demonstrates the need for more research and the creation of a new dataset that incorporates realistic networks involving multiple Windows operating systems, IoT and Industrial IoT (IIoT) services. More importantly, the existing datasets focus entirely on process system calls, ignoring other sources of traces like the memory, hard drive or the processor state. Furthermore, there is a lack of IoT-derived data representation in the existing datasets. This paper seeks to address these shortcomings by proposing a new Windows dataset from a new IoT testbed network designed at the IoT lab of the University of New South Wales, Canberra.

\section{Proposed Testbed for generating federated Windows TON\_IoT datasets}
\label{testbed}

\begin{figure*}[t]
\centering
\includegraphics[width=13cm,height=12cm]{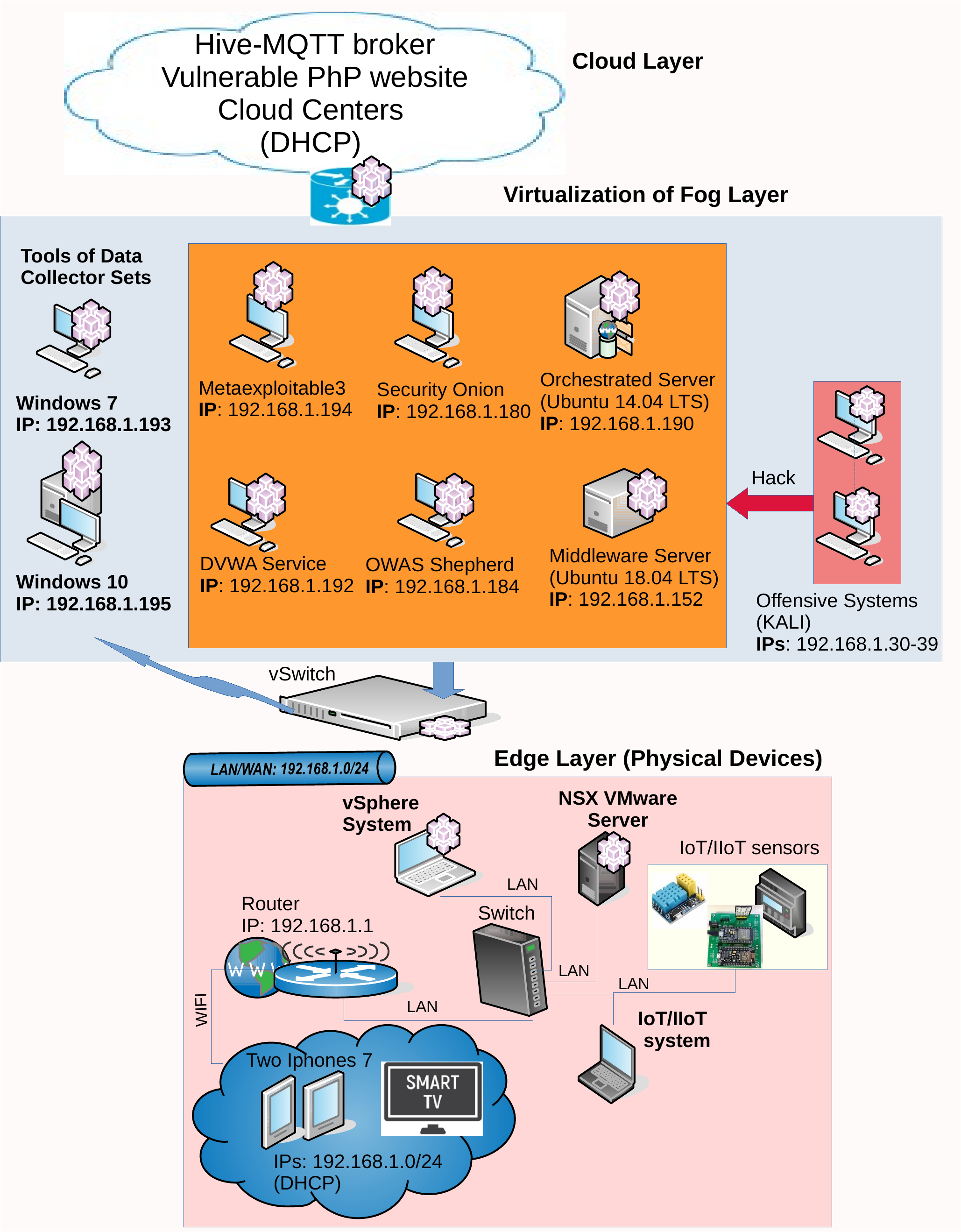}
\caption{Configured testbed of TON\_IoT datasets for collecting Window data }
\label{fig1}
\end{figure*}

The proposed testbed architecture of the ToN\_IoT datasets for gathering audit traces of Windows operating systems is shown in Figure \ref{fig1}. The testbed was designed based on interacting network and IoT systems with the three layers of edge, fog and cloud to mimic the realistic implementation of recent real-world IoT networks. The dynamics of the three layers, involving physical and simulated systems, were flexibly managed by the technologies of SDN and NVF. The NSX-VMware data center platform \cite{SDN} was used to provide an SDN solution for the proposed testbed of the TON\_IoT datasets. This technology permits the creation of overlay networks with the same capabilities of physical networks.

The VMware NSX platform was deployed with the VMware NFV hypervisor to allow the creation and management of various virtual machines that concurrently operate to offer the IoT and network services. In VMware NSX, the vCloud NFV platform was employed to provide a modular design with abstractions that enable multi-domain, hybrid physical and VM deployments \cite{NFV}. The NSX vCloud NFV platform enables the design of a dynamic testbed IoT network of the ToN\_IoT with creating and controlling several Virtual Machines (VMs) for hacking and normal operations, allowing the communications between the edge, fog and cloud layers.

\textbf{The components of the testbed are explained for the three layers as follows: }

\begin{enumerate}

\item \textbf{Edge layer} - involves the physical devices and their operating
systems utilized as the infrastructure of configuring the virtualization
technology and cloud services at the layers of fog and cloud, respectively.
It includes multiple IoT/IIoT devices, such as Modbus, light bulb
sensors, smartphones, smart TVs, as well as host systems, such
as workstations and servers, used to connect IoT/IIoT devices, hypervisors
and physical gateways (i.e., routers and switches) to the Internet.
The hypervisor technology of NSX-VMware was installed on a host server
at the edge layer to manage the Virtual Machines (VMs) created at
the fog layer. 

\item \textbf{Fog layer}- includes the virtualization technology that controls
the VMs and their services using the NSX-VMware and vCloud platform
to offers the framework of executing SDN and NFV in the proposed testbed.
The NSX vCloud NFV platform enables the design of a dynamic testbed
IoT/IIoT network of the ToN\_IoT with creating and controlling several
VMs for hacking and normal operations, allowing the communications
between the edge, fog and cloud layers. This layer includes the nodes
of virtual machines configured to generate the datasets, as explained
in the following: 

\begin{itemize}

\item \textit{\textbf{ Orchestrated server}} - is one of the main virtualized servers configured in the testbed using the Ubuntu 14.04 LTS with the IP address (\textit{192.168.1.190}). This server offered many orchestrated services, such as FTP, Kerberos,
HTTPS, and DNS to simulate real production networks and generate more
simulated network traffic using the Ostinato Traffic Generator \cite{Ostinato}
that transmits traffic to other VMs in the testbed.

\item  \textit{\textbf{Middleware server}}- is the IoT/IIoT virtualized server deployed in
the testbed using the Ubuntu 18.04 with the IP address (\textit{192.168.1.152}).
This server included the scripts that run IoT/IIoT services through
public and local MQTT gateways utilized in the testbed and linked
with the cloud layer to subscribe and publish the telemetry data of
IoT/IIoT sensors. 

\item \textit{\textbf{Client Systems}} - include a Windows 7 VM (IP address: \textit{192.168.1.193}),
Windows 10 VM (192.168.1.195), DVWA web service (\textit{192.168.1.192}),
OWASP security Sphered VM (192.168.1.184), Metaspoitable 3 (\textit{192.168.\\1.194}). The two windows were used as the remote web interface
of the node-red IP (\textit{192.168.1.152}) and their network traffic and audit
traces were logged. The Damn Vulnerable Web App (DVWA) \cite{DVWA} was
utilized to make security vulnerabilities through web applications
hacked using the virtualized offensive systems. The OWASP security
Sphered VM \cite{Shepherd} is an open-source platform that has many security
vulnerabilities against mobile and web applications exploited using
the offensive systems. In addition, the Metasploitable3 VM \cite{Metasploitable3}
was deployed in the testbed to increase vulnerable fog nodes and hack
them using various attacking techniques by the offensive systems. 

\item \textit{\textbf{Offensive systems}} - include the kali Linx VMs and scripts of hacking
scenarios that exploit vulnerable systems in the testbed network.
Ten static IP addresses (i.e.,\textit{192.168.1.30-39}) were employed in the
testbed to launch attacking scenarios and breach vulnerable systems
either IoT/IIoT services (client and public MQTT brokers and node-red
IP), operating systems (i.e., Windows 7 and 10, and Ubuntu 14.04 LTS
and 18.04 LTS), and network systems (i.e., IP addresses and open protocols
of the VMs).

\item \textit{\textbf{Data Logger Systems}} - log audit traces of Windows 7 and 10
operating systems included in the testbed. The Data Collector Set
tool arranges data collection points, such as performance counters
and event trace data, into a single collection \cite{windows-collectors}. Data Collector
Sets enable us to schedule data collection so that the data can be
anaysled and generated in CSV files, as in our datasets. While launching
normal and hacking scenarios, the tools of Data Collector Sets in
both Windows VMs were automatically configured to log data features
of memories, networks, hard disks, processors and processes that happened. Data that
were collected for performance counters by the tools of Data Collector
Sets configured on Windows 7 and Windows 10 were stored in log files
and can be opened using the Windows Performance Monitor tool (\textit{.blg format}) \cite{windows-monitor}. The data features from generated for the Windows TON\_IoT
dataset are discussed below. 

\end{itemize}
\item \textbf{Cloud layer}- contains the cloud services configured online
in the testbed, as shown in Figure \ref{fig1}. The fog and edge services connected
with the public HIVE MQTT dashboard \cite{MQTT}, a public PHP vulnerable
website \cite{PhP-vulnerable}, cloud virtualization, and cloud data analytics services
(e.g., Microsoft Azure or AWS). The public HIV MQTT dashboard enabled
us to publish and subscribe to the telemetry data of IoT/IIoT services
via the configuration of the node-red tool. The public PHP vulnerable
website used to launch injection hacking events against websites.
The other cloud services were configured either in Microsoft Azure
or AWS to transmit sensory data to the cloud and visualize their patterns. 
\end{enumerate}

\section{Hacking Techniques launched in Testbed}  
\label{hacking}

Hacking scenarios were utilized to launch nine attack categories against vulnerable elements of IoT/IIoT applications, operating systems and network systems. The scripts and some links of the attacking categories have been published in \cite{ton-data}. The nine attack families employed in the datasets are explained as follows: 
\begin{enumerate}

 \item\textbf{Scanning attack} - We used the Nessus and Nmap tools from the offensive systems with IP addresses (\textit{192.168.1.20-38}) against the target subnet (\textit{192.168.1.0/24}) and all other public vulnerable systems such as the Public MQTT broker and vulnerable PHP website.  For example, nmap \textit{192.168.1.40-254}, and the scans of the Nessus tool for the same range of IP addresses. 
 
 \item \textbf{Denial of Service (DoS) attack} - We utilized DoS attack scenarios on the offensive systems with IP addresses (\textit{192.168.1.{30,31,39}}) to hack vulnerable elements in the IoT testbed network. We created Python scripts using the Scapy package to launch the DoS attacks. 
 
 \item \textbf{Distributed Denial of Service (DDoS) attack} - We used DDoS attacks in the offensive systems wit IP addresses (\textit{192.168.1.{30,31,34,35,36,37,38}}) to breach several weaknesses in the IoT testbed network. We developed Python scripts using the Scapy package to launch the DoS attacks. Further, automated bash scripts were developed to launch DDoS against vulnerable nodes of the testbed using the ufonet toolkit.   
 
 \item \textbf{Ransomware attack} - We used the Kali Linux with IP addresses (\textit{192.168.1.{33, 37}}) to execute this malware against windows operating systems and their webpages of monitoring IoT services included in the testbed network. This attack executed using the Metasploit framework that hacks the SMB vulnerability of the systems, named eternalblue.  
 
 \item \textbf{Backdoor attack} - We used the offensive systems with IP addresses (\textit{192.168.1.{33,37}}) to keep the hacking persistence using the Metasploit framework by executing a bash script of the command “run persistence -h”. 
 
 \item \textbf{Injection attack} - We used various injection scenarios from the offensive systems with IP addresses (\textit{192.168.1.{30, 31, 33, 35}}) to inject data inputs against web applications of DVWA and Security Shepherd VMs and webpages of IoT services through other VMs, including SQL injection, client-side injection, broken authentication and data management, and unintended data leakage.

 \item \textbf{Cross-site Scripting (XSS) attack} -  We employed the offensive systems with IP addresses (\textit{192.168.1.{32,35,36,39}}) to illegally inject web applications of DVWA and Security Shepherd VMs and webpages of IoT services through other VMs. In these systems, we created malicious bash scripts of python codes to hack the web applications of the testbed network using the Cross-Site Scripter toolkit (named XSSer).
 
 \item \textbf{Password attack} - We used the offensive systems with IP addresses (\textit{192.168.1.{30, 31, 32, 35, 38}}). In these systems, the hydra \cite{Hydra} and cewl toolkits were configured using automated bash scripts to concurrently launch password hacking scenarios against vulnerable nodes in the testbed.
 
 \item \textbf{Man-In-The-Middle (MITM) attack} - We utilized the offensive systems with IP addresses (\textit{192.168.1.{31,34}}) to launch various MITM scenarios in the testbed network. In the systems,  we employed the Ettercap tool to execute ARP spoofing, ICMP redirection, port stealing and DHCP spoofing.  
\end{enumerate}

\section{Experimental Results and Discussions}
\label{expereiments}
\subsection{Feature Generation and Labelling}

The Windows datasets were generated using the virtual machines running Windows 7 and Windows 10 and incorporated the collections of data from multiple sources, including memory, process, processor and hard drive of the systems. In order to generate the Windows datasets, the collectors of the Performance Monitor Tool \cite{windows-monitor} were executed on each machine. The initial raw version of the datasets was collected in a \textit{blg} format, and through the Performance Monitor tool, the disk, process, processor, memory activities were extracted and saved in a \textit{CSV} format. As explained in \cite{ton-data}, the Windows 7 dataset generated 133 features and other two attributes of the class label, either normal or attack, and attack types of the nine attacked used in the testbed. The Windows 10 dataset generated 125 features and the other two attributes of the class label and attack types.

After generating the data features of Windows 7 and Windows 10, we added the two attributes for labeling each record either normal or attack type. The labeling process was used by the ground truth CSV files that contain the attack events that took place while running the testbed. The timestamp ‘\textit{ts}’ attribute was matched against each record in the CSV files, where if the ‘\textit{ts}’ of the ground truth table equals the ‘\textit{ts}’ of the data records in the CSV files, the records were labeled as attacks; otherwise, they were labeled as normal. The authentic labeling process of the datasets proves the fidelity of the correct security events that occurred during the implementation of the testbed and its authenticity for evaluating cyber security solutions based on machine learning algorithms.

\subsection{Statistics of TON\_IoT Windows datasets}

There is a huge number of normal and attack types collected in the Windows 7 and Windows 10 datasets, as listed in Table \ref{table1}. It can be seen that the windows 7 dataset includes 28367 records while the Windows 10 dataset involves 35975 records of both normal and attack observations. These datasets include the CSV files of training-testing sets for training and testing machine learning models, as published in \cite{ton-data}.  

\begin{table} 
\centering
\caption{Number of normal records and each type of attach collected in the Windows 7 and 10 datasets} \label{table1}
\begin{tabular}{|c| c| c|}
\hline 
\textbf{Type of  events} & \textbf{Windows 7} & \textbf{Windows 10}\\
\hline 
Normal & 22387 & 24871\\
\hline 
Backdoor & 1779 & -\\
\hline 
DDoS & 2134 & 4608\\
\hline 
Ransomware & 82 & -\\
\hline 
Injection & 998 & 612\\
\hline 
XSS & 4 & 1268\\
\hline 
Password & 757 & 3628\\
\hline 
Scanning & 226 & 447\\
\hline 
DoS & - & 525\\
\hline 
MITM & - & 15\\
\hline 
\end{tabular}
\end{table}

The training-testing set of Windows 7 involves 10,000 normal records and 5980 attack records, whilst the training-testing set of Windows 10 contains  10,000 normal records and 11104 attack records, as demonstrated in Table \ref{table1}. It is observed that all the attack types included in the entire sets are used in Table \ref{table1}. The training-testing sets were carefully analyzed to include different attack types and normal behaviors included in the entire sets. \textbf{Researchers can then divide these sets for two subsets, such as 70\% for training and 30\% for testing or k-fold cross-validation models, for testing the performances of AI-enabled cyber security models.  }

\begin{table}
\centering
\caption{10 most correlated features (their data types are numeric) in the Windows 7 dataset. }
\label{win7-features}
\begin{tabular}{| p{2.5cm} | p{5.5cm} |}
\hline 
 \textbf{Feature name} & \textbf{Feature Description}\\
\hline 
 Process\_Total\_IO Other\_Bytes\_sec & The process issues bytes to I/O operations that do not involve data such as control operations.\\
\hline 
 Network\_I.Intel\_R Pro\_1000MT\_Bytes Received\_sec & The rate at which bytes are received over each network adapter, including framing characters.\\
\hline 
 Process\_Total\_IO Other Operations\_sec & The rate at which the process  issues I/O operations that are neither
read nor write operations (for example, a control function). \\
\hline 
 Process\_Total\_IO Data Bytes\_sec & The rate at which the process is reading and writing bytes in I/O
operations.\\
\hline 
 Process\_Total\_IO Read Bytes\_sec & The rate at which the process is reading bytes from I/O operations.\\
\hline 
 Network\_I.Intel.R.\_ Pro\_1000MT\_Bytes Received\_sec & The rate at which bytes are received over each network adapter. \\
\hline 
 Process\_Pool\_Paged\_ Bytes & The size, in bytes, of the paged pool, an area of the system's virtual memory that is used for objects that can be written to disk when they
are not being used. \\
\hline 
 Network\_I.Intel.R\_Pro\_ 1000MT.Bytes\_Sent\_ sec & The rate at which bytes are sent over each network adapter, including
framing characters. \\
 \hline 
 Network\_I.Intel.R\_Pro\_ 1000MT.Packets Received\_sec & The rate at which packets are received on the network interface.\\
\hline 
 Process.Total\_IO Data Operations\_sec & The rate at which the process is issuing read and write I/O operations.\\
\hline 
\end{tabular}
\end{table}

\begin{table}
\caption{10 most correlated features (their data types are numeric) in the Windows 10 dataset. }
\label{win10-features}
\begin{tabular}{ | p{2.5cm} | p{5.5cm} |}
\hline 
\textbf{Feature name} & \textbf{Feature Description}\\
\hline 
 Network\_I.Intel.R\_ 82574L\_GNC.Current Bandwidth & The current bandwidth of the network interface in bits per second
(BPS). \\
\hline 
 Network\_I.Intel.R\_ 82574L\_GNC.Packets Sent Unicast.sec & The rate at which packets are requested to be transmitted to subnet-unicast addresses by higher-level protocols.\\
\hline 
 Memory.Pool.Paged Bytes & The size, in bytes, of the portion of the paged pool that is currently resident and active in physical memory. \\
\hline 
 LogicalDisk.\_Total Disk Read Bytes.sec & The rate at which bytes are transferred from the disk during reading
operations.\\
\hline 
 Memory.Page Reads.sec & The rate at which the disk was read to resolve hard page faults. \\
\hline 
 Network\_I.Intel.R\_ 82574L\_GNC.Packets Sent.sec & The rate at which packets are sent on the network interface.\\
\hline 
 Memory.Modified. Page List Bytes & The amount of physical memory, in bytes, that is assigned to the modified
page list. \\
\hline 
 Process\_IO Data Operations\_sec & The rate at which the process is issuing read and write I/O operations. \\
\hline 
 LogicalDisk\_Total. Avg.Disk.Bytes Transfer & The average number of bytes transferred to or from the disk during
write or read operations.\\
\hline 
 Processor\_pct\_ Processor\_Time & The amount of elapsed time that the processor spends executing a non-Idle thread. \\
\hline 
\end{tabular}
\end{table}

\subsection{Correlation Analysis of Features}

\begin{figure}
\includegraphics[width=9cm,height=10cm]{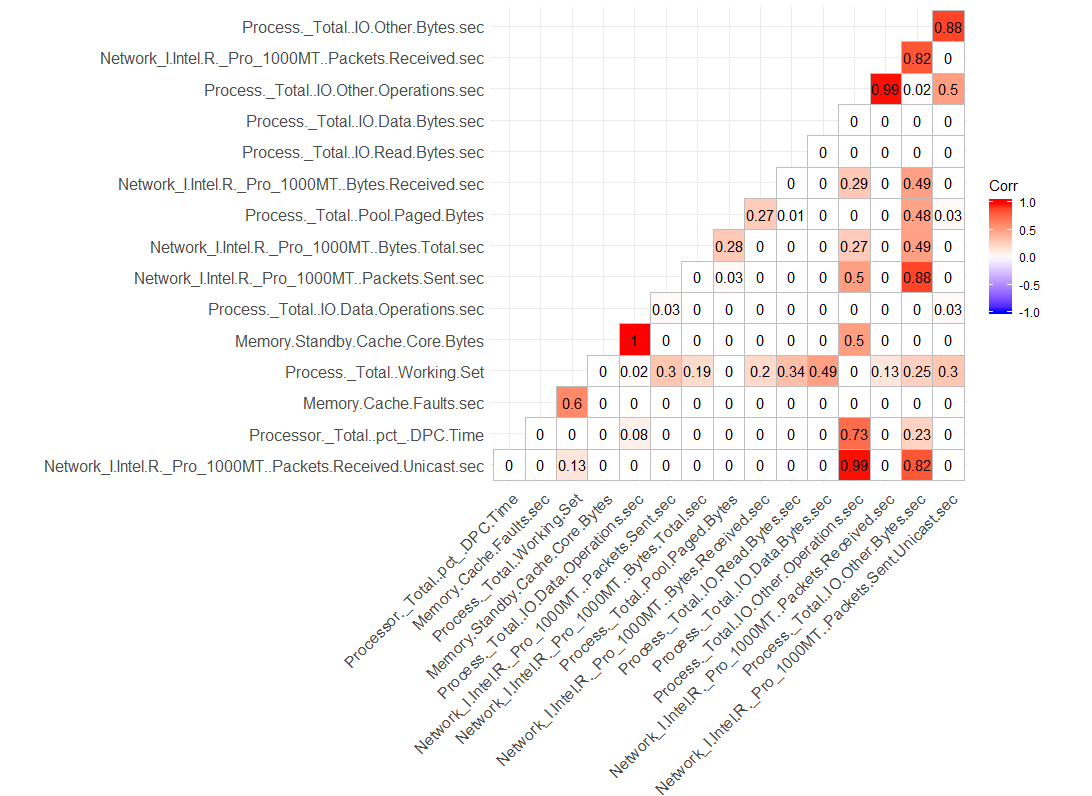}
\caption{Correlation matrix of most important features in the Windows 7 dataset} \label{fig2}
\end{figure}

\begin{figure}
\includegraphics[width=9cm,height=10cm]{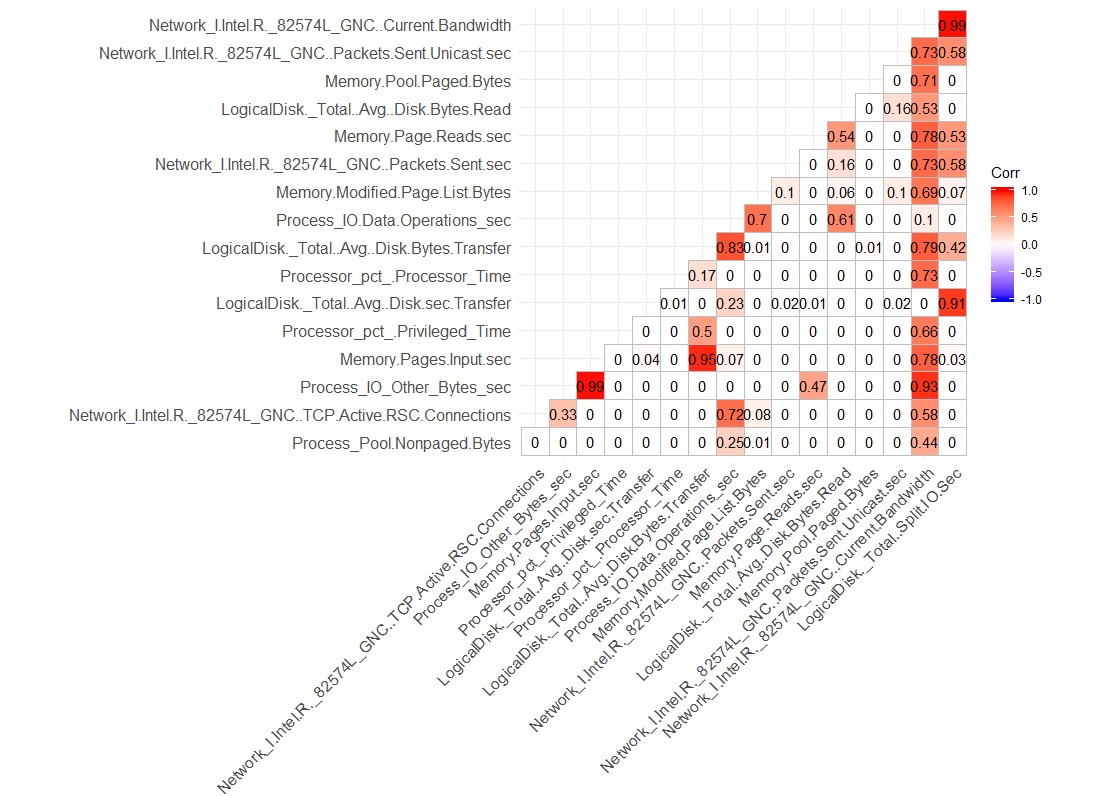}
\caption{Correlation matrix of most important features in the Windows 10 dataset} \label{fig3}
\end{figure}

The correlation analysis has a great impact to demonstrate the strength of features and their utility to define security events using machine learning models. In order to estimate the correlation coefficient between the features without the label attributes on the Windows 7 and 10 datasets, we developed a correlation coefficient function \cite{koroniotis2019towards} in the R programming language to rank the features’ strengths into a range of [-1, 1].The sign of the correlation coefficient refers to the direction of the relationship, while the magnitude of the correlation (i.e, how close it is to -1 or +1) indicates the strength of the relations between the features \cite{koroniotis2019towards}.

The correlation matrix was adapted to select the most correlated features with higher than or equal a cut-off value of 0.85\%. The most 10 correlated features in both datasets are described in Tables \ref{win7-features} and \ref{win10-features}. The description of the rest features can be found in \cite{ton-data}. The most correlated features of the windows 7 and 10 datasets are shown in Figures \ref{fig2} and \ref{fig3}, respectively.  The most correlated features would be used for training and validating machine/deep learning algorithms to evaluate their efficiency in classifying attack families included in the datasets.  


\section{Conclusion}
\label{conclusion}
This paper has introduced the description and preliminary results of the Windows TON\_IoT  datasets created at the IoT lab of UNSW Canberra. In order to create the federated datasets, a new IoT testbed  was designed that included a wide variety of IoT services deployed at the edge layer, virtual machines of operating systems configured at the fog layer, as well as cloud services constructed at the cloud layer. The dynamic interaction between the three layers was deployed using the VMware NSX and vCloud NFV platforms to provide SDN and NVF services. Recent normal and nine attack categories were executed in the datasets in order to generate authentic data sources for assessing the reliability of new AI-based cyber security applications. Further, the features of Windows 7 and Windows 10 data were collected from the audit traces of memories, processors, processes and hard disks to ensure the identification of new attack patterns that could stealthy exploit windows operating systems. A large number of data samples was collected for windows 7 and windows 10 datasets. The collected datasets show a wide variety of normal and attack events, revealing the fidelity of the datasets for assessing new AI-based cyber security applications, including intrusion detection, privacy preservation, digital forensics, as well as threat intelligence and hunting, in which we will explore in the future.



\end{document}